\documentclass[a4paper,11pt]{article}
\usepackage{jcappub} % for details on the use of the package, please
\usepackage[T1]{fontenc}
\usepackage{amsmath}
\usepackage{amsfonts}
\usepackage{amssymb}
\usepackage{mathrsfs}

\usepackage{graphicx}
\usepackage{color}
\usepackage{soul}
\usepackage{subfig}
\usepackage[normalem]{ulem}
\usepackage[utf8]{inputenc}
\usepackage{aas_macros}
\usepackage[T1]{fontenc} % if needed

\renewcommand{\vec}[1]{\boldsymbol{#1}}

\usepackage{floatrow}
\DeclareFloatFont{tiny}{\scriptsize}
\title{\boldmath Possible bump structure of cosmic ray electrons unveiled by AMS-02 data and its common origin along with the nuclei and positron}

\author[a,b]{Pei-pei Zhang,}
\author[b,1]{Bing-qiang Qiao,\note{Corresponding author.}}
\author[b]{Wei Liu,}
\author[a,1]{Shu-wang Cui,}
\author[c,d]{Qiang Yuan}
\author[b,e]{ and Yi-qing Guo}

\affiliation[a]{Hebei Normal University, Shijiazhuang 050024 , Hebei, China}
\affiliation[b]{Key Laboratory of Particle Astrophysics, Institute of High Energy Physics, Chinese Academy of Sciences, Beijing 100049, China}
\affiliation[c]{Key Laboratory of Dark Matter and Space Astronomy, Purple Mountain Observatory, Chinese Academy of Sciences, Nanjing 210008, China}
\affiliation[d]{School of Astronomy and Space Science, University of Science and Technology of China, Hefei 230026, China}
\affiliation[e]{University of Chinese Academy of Sciences, Beijing 100049, China}

\emailAdd{qiaobq@ihep.ac.cn}
\emailAdd{cuisw@hebtu.edu.cn}

\abstract{The local pulsar and its progenitor, SNR, can together accelerate the positron, electron and nuclei to very high energy. The famous excesses of positron(nuclei) above $20$($200$) GeV possibly come from such kind of local source. This hints that the primary electron should also hold "excess" above $200$ GeV, synchronously accelerated along with the nuclei. The recent precise measurement of sharp dropoff at 284 GeV of positron by AMS-02 experiment takes chance to study this expected electron excess. In this work, the spatially-dependent propagation with a local source is used to reproduce the spectrum of positron, electron and proton. When considering the dropoff at 284 GeV of positron, a sharp bump structure for primary electron above 284 GeV is required to fit the total spectrum of positron and electron. Then we systematically study the common origin of the excesses of positron, electron and nuclei from Geminga pulsar and SNR. Those excesses can be reproduced under this unified single-source model. Lastly, we hope that the fine bump structure can be observed to support our model by AMS-02 experiment in future.}

\begin{document}
\maketitle
\flushbottom

\section{Introduction}
\label{sec:intro}

In recent decades, great progress has been made in development of advanced space-borne and ground-based experiments. With those new generation of experiments, comic rays (CRs) measurements are stepping into an era of high precision and revealing a series of new phenomena. A fine structure of spectral hardening above 200 GV for nuclei were observed by ATIC-2, CREAM and PAMELA experiments \citep{2007BRASP..71..494P,2009BRASP..73..564P,2010ApJ...714L..89A,2017ApJ...839....5Y,2011Sci...332...69A}. AMS-02 experiment confirmed this discovery with full accuracy \citep{2015PhRvL.114q1103A,2015PhRvL.115u1101A}. {Just recently, more interesting was that the break-off at 14 TeV for proton was clearly observed by CREAM, NUCLEON and DAMPE experiment \citep{2017ApJ...839....5Y,2017JCAP...07..020A,2018JETPL.108....5A,2019SciA....5.3793A}.} Several kinds of models have been proposed to explain
the origin of spectral hardening, which include the local source \citep{2013APh....50...33S,2017PhRvD..96b3006L,2019JCAP...10..010L,2012ApJ...752...68V}, the combined effects from different group sources and the spatially-dependent propagation \citep{2016ApJ...819...54G,2016ChPhC..40a5101J,2018PhRvD..97f3008G,2018ApJ...869..176L,2020arXiv201002826M}. Considering the break-off at 14 TeV in spectrum, it seems that the local single source model becomes natural and accessible. But where is the dominant source ? Further clues of observation is necessary to support this view of point.

The anisotropy of CRs is an optimal choice to fulfill this role. Thanks to new generation of ground-based experiments, the measurements of anisotropy has made great progress from hundred of GeV to several PeV \citep{2006Sci...314..439A,2017ApJ...836..153A,2016ApJ...826..220A,2013ApJ...765...55A,2015ApJ...809...90B}. It is clear that the phase of anisotropy below 100 TeV roughly directs to galactic anticenter, which is totally paradoxical with the traditional propagation model (TPM). However, above 100 TeV, the phase directs to galactic center and is consistent with the expection of TPM. Correspondingly, the amplitude has experienced similar transition at the critical energy of 100 TeV. In addition, the most importance is that there exists a common transition energy scale between the structures of the energy spectra and the anisotropy. The local source possibly plays very important role to resolve the conjunct problems of spectra and anisotropy. Furthermore, the direction of anisotropy can roughly outline the position of such local source. In our recent work, we propose a local source under the spatially-dependent propagation (SDP) to reproduce the co-evolution of the spectra and anisotropy. In this model, the optimal candidate of local source is SNR at Geminga's birth place \citep{2019JCAP...10..010L,2019JCAP...12..007Q}. If this standpoint is right, one question is that whether the electron has similar excess above 200 GeV, synchronally accelerated along with the nuclei. In addition, can the Geminga pulsar dominantly contribute the positron spectrum?

The famous spectral excess of positron above $20$ GeV was discovered by PAMELA experiment \citep{2009Natur.458..607A, 2014PhRvL.113l1101A}. Then the AMS-02 experiment confirmed this remarkable result with unprecedented precision \citep{2019PhRvL.122d1102A}. Just recently, a sharp dropoff at 284 GeV was observed by AMS-02 experiment
with above 4$\sigma$ confidence level \citep{2019PhRvL.122d1102A}. However, the dropoff around 1 TeV in the total spectrum of positron and electron was first reported by HESS collaboration \citep{2008PhRvL.101z1104A,2009A&A...508..561A, hess-icrc2017} and validated by MAGIC \citep{2011ICRC....6...47B}, VERITAS \citep{2015arXiv151001269S} experiments. DAMPE experiment performed the first direct measurement to this feature and announced that the breakoff was at $\sim$0.9 TeV \citep{2017Natur.552...63D}. At this erengy range, CALET experiment also gave the corresponding messurement \citep{2018PhRvL.120z1102A}. It means that the total spectrum shows a single power-law from $\sim$40 GeV to $\sim$0.9 TeV without significant features \citep{2019PhRvL.122j1101A,2017Natur.552...63D}. Considering the lower energy cutoff at 284 GeV, this deficit of positron spectrum above 284 GeV should be supplied by the primary electron, which indicates that an extra bump structure is required for the primary one above $\sim$284 GeV. It’s worth noting that similar origins are predicted in \citep{2013PhLB..727....1Y, 2012PhRvL.108u1102E, 2013PhRvD..88b3013C, 2015PhLB..749..267L, 2020arXiv200715321F}.
A number of alternatives have been proposed, which can be either astrophysical, such as local pulsars and the hadronic interactions inside SNRs, or even more exotic dark matter self-annihilation or decay \citep{2009PhRvL.103e1101Y, 2009A&A...497...17V, 2009PhRvD..80f3003F, 2009PhRvL.103c1103B, 2009PhRvD..79b3512Y, 2009MPLA...24.2139H, 2009ApJ...700L.170H, 2010SCPMA..53..842W, 2010IJMPD..19.2011F, 2012APh....39....2S, 2012Prama..79.1021C, 2013FrPhy...8..794B, 2016PTEP.2016b1E01K, 2017PhRvD..96b3006L}. Among those models, the local pulsar source \citep{2011JCAP...02..031M, 2012A&A...544A..92B,2012APh....39....2S, 2014JCAP...04..006D}, such as Geminga pulsar, is favourite candidate \citep{2018ApJ...863...30F, 2019MNRAS.484.3491T}, though there leaves some debates \citep{2017Sci...358..911A}.

Based on above discussion, a sharp bump structure possibly exist in the spectrum of primary electron. Geminga SNR and pulsar are the optical candidate to supply those anomalous features in spectra and anisotropy. Therefore, in this work we firstly deduce the possible bump structure based on the different break-off of positron and electron. Then a common origin from Geminga pulsar and SNR was proposed to explain the positron excess and the spectral hardening of proton/electron starting from 20 and 200 GeV respectively. The paper is organized as follows. Sec.2 describes the SPD model and Geminga source briefly, Sec.3 presents the calculated results and Sec.4 gives the conclusion.

\section{Model Description}

\subsection{Spatially-dependent propagation}

The SDP of CRs has received a lot of attention in recent years. It was first introduced as a Two Halo model (THM) \citep{2012ApJ...752L..13T} to explain the spectral hardening of both proton and helium  above $200$ GeV \citep{2011Sci...332...69A}. Afterwards, it is further applied to secondary and heavier components \citep{2015PhRvD..92h1301T, 2016PhRvD..94l3007F, 2016ApJ...819...54G, 2018ApJ...869..176L, 2020ChPhC..44h5102T, 2020arXiv200701768Y}, diffuse gamma-ray distribution \citep{2018PhRvD..97f3008G} and large-scale anisotropy \citep{2019JCAP...10..010L, 2019JCAP...12..007Q}. For a comprehensive introduction, one can refer to \cite{2016ApJ...819...54G} and \cite{2018ApJ...869..176L}.

In the SDP model, the whole diffusive halo is divided into two parts. The Galactic disk and its surrounding area are called the inner halo (IH) region, in which the diffusion coefficient is spatially-dependent and relevant to the radial distribution of background CR sources. The extensive diffusive region outside the IH is named as the outer halo (OH) region, where the diffusion is regarded as only rigidity dependent. The spatially-dependent diffusion coefficient $D_{xx}$ is thus parameterized as:
\begin{equation}
D_{xx}(r,z, {\cal R} )= D_{0}F(r,z)\beta^{\eta} \left(\dfrac{\cal R}{{\cal R}_{0}^{'}} \right)^{\delta(r,z)} ~,
\label{eq:diffusion}
\end{equation}

For the parameterization of $F(r,z)$ and $\delta(r,z)$, one can refer to \cite{2020ChPhC..44h5102T}.
The size of IH is represented by its half thickness $\xi z_{h}$, whereas the OH region's is $(1-\xi) z_{h}$.

In this work, we adopt the common diffusion-reacceleration (DR) model, with the diffusive-reacceleration coefficient $D_{pp}$ coupled to $D_{xx}$ by $D_{pp}D_{xx} = \dfrac{4p^{2}v_{A}^{2}}{3\delta(4-\delta^{2})(4-\delta)}$, in which $v_A$ is the so-called Alfv\'en velocity \citep{1994ApJ...431..705S}. The numerical package DRAGON is used to solve the SDP equation to obtain the distribution of comic ray electrons (CREs). During the propagation, the energetic CREs still suffer from the energy loss from synchrotron radiation and inverse Compton scattering \citep{1998ApJ...509..212S, 1998ApJ...493..694M}. A full-relativistic treatment of the inverse-Compton losses \citep{2010A&A...524A..51D} has been implemented in the DRAGON package \citep{2017JCAP...02..015E}. Less than tens of GeV, the CR fluxes are impacted by the solar modulation. The well-known force-field approximation \citep{1968ApJ...154.1011G, 1987A&A...184..119P} is applied to describe such an effect, with a modulation potential $\phi$ adjusted to fit the low energy data.

\subsection{Supernova remnants}

The SNRs are regarded as the most likely sites for the acceleration of CRs by default, in which the charge particles are accelerated to a power-law distribution through the diffusive shock acceleration. The distribution of SNRs are approximated as an axisymmetric, which is usually parameterized as
\begin{equation}
f(r, z) = \left(\dfrac{r}{r_\odot} \right)^\alpha \exp \left[-\dfrac{\beta(r-r_\odot)}{r_\odot} \right] \exp \left(-\dfrac{|z|}{z_s} \right) ~,
\label{eq:radial_dis}
\end{equation}
where $r_\odot \equiv 8.5$ kpc represents the solar distance to the Galactic center. The parameters $\alpha$ and $\beta$ are taken as $1.09$ and $3.87$ respectively in this work \citep{2015MNRAS.454.1517G}. Perpendicular to the Galactic plane, the density of CR sources descends exponentially, with a mean value $z_{s} = 100$ pc. The axisymmetric approximation is plausible as the diffusion length of CR nuclei is usually much longer than the characteristic spacing between the neighbouring spiral arms. However, subject to the energy loss from synchrotron radiation and inverse Compton scattering, the transport distance of the energetic electrons is much shorter so that the above approximation may no longer hold. Thus the inclusion of a more realistic description of the source distribution is expected to have striking impact on the observed spectrum of high-energy electrons.

As is well-known, our Milky Way is a spiral galaxy. In this work, the spiral distribution of SNRs follows the model established by \cite{2006ApJ...643..332F}. The Galaxy consists of four major arms, with the locus of the $i$-th arm centroid expressed as a logarithmic curve: $\theta(r) = k^{i} \ln(r/r^{i}_{0}) + \theta_{0}^i$, where $r$ is the distance to the Galactic center. For the values of $k^{i}$, $r^{i}_{0}$ and $\theta_{0}^i$ for each arm, one can refer to \cite{2020ChPhC..44h5102T}. Along each spiral arm, there is a spread in the normal direction which follows a Gaussian distribution, i.e.
\begin{equation}
f_i = \dfrac{1}{\sqrt{2\pi} \sigma_i} \exp \left[-\dfrac{(r-r_i)^2}{2\sigma^2_i} \right] ~, ~~~ i \in [1,2,3,4] ~,
\end{equation}
where $r_i$ is the inverse function of the $i$-th spiral arm's locus and $\sigma_i$ is equal to $0.07 r_i$. The number density of SNRs at different radii still conforms with the radial distribution in the axisymmetric case, i.e. Eq (\ref{eq:radial_dis}). The injection spectra of nuclei and electron are assumed to have a broken power-law, i.e.
\begin{equation}\label{eq:spectrum_CRE}
\mathcal{Q}({\cal R}) = \mathcal{Q}_{0} \left\{
\begin{array}{lll}
\left(\dfrac{{\cal R}_{\rm br1} }{{\cal R}_{0}} \right)^{\nu_{2}}
\left(\dfrac{{\cal R}}{{\cal R}_{\rm br1}} \right)^{\nu_{1}},  & {{\cal R} \leqslant {\cal R}_{\rm br1}} \\
\\
\left(\dfrac{{\cal R}}{{\cal R}_{0}} \right)^{\nu_{2}},  & {\cal R}_{\rm br1} < {\cal R} \leqslant {\cal R}_{\rm br2} \\
\\
\left(\dfrac{{\cal R}_{\rm br2} }{{\cal R}_{0}} \right)^{\nu_{2}} \left(\dfrac{{\cal R}}{{\cal R}_{\rm br2}} \right)^{\nu_{3}},  & {\cal R} > {\cal R}_{\rm br2}
\end{array}
\right.
\end{equation}
%\textcolor {red}
 {where $\mathcal{Q}_0$, $\nu_{1/2/3}$, $R_{0}$ and $R_{\rm br1/2}$ are the normalization, power index, reference rigidity and broken rigidity respectively.}

\subsection{Local source: Geminga}

The Geminga SNR locates at its birth place with the distance and age of $r = 330$ pc and $t_{\rm inj} = 3.3\times 10^{5}$ yrs \citep{1994A&A...281L..41S}. 
 {Moreover, the observed direction is roughly consistent with the large scale anisotropy of CRs below $100$ TeV. However some authors think that the large anisotroy was caused by the local magnetic field and has no relation to the direction of astrophysical sources \citep{2016PhRvL.117o1103A,2019JPhCS1181a2039K,2020ApJ...892....6L}.}
The current Geminga pulsar is located at $250$ pc to solar system \citep{2007Ap&SS.308..225F}. In this work, we assume that Geminga pulsar accounts for the positron excess above $\sim 20$ GeV and Geminga SNR is responsible for the excesses of the electron and nuclei above $200$ GeV.

The propagation of positrons, electrons and nuclei from Geminga SNR and pulsar is described by the time-dependent propagation equation, i.e.
\begin{equation}
\frac{\partial \varphi}{\partial t} -D_{xx}\Delta \varphi +\frac{\partial}{\partial E}(\dot{E} \varphi) = Q(E, t) \delta(\vec{r}-\vec{r}^{\prime}) ~.
\label{eq:t_dep_diffu_e}
\end{equation}
For the nuclei, the energy loss can be safely neglected, i.e. $\dot{E} = 0$. But the energy loss of the electrons and positrons is significant and has to be taken into account during their propagation. Here their energy loss rate $\dot{E}$ is approximated as $\dot{E} = -bE^2$ in the Thomson limit\cite {1995PhRvD..52.3265A}. They are injected into the  interstellar space as a power-law spectrum plus a high energy cutoff, i.e.
\begin{equation}
Q({\cal R}, t)=Q_{0}(t) \left(\dfrac{\cal R}{1 ~\rm GV}\right)^{-\gamma} \exp \left[-\dfrac{\cal R}{{\cal R}_{\rm c} } \right] ~,
\label{eq:nearby}
\end{equation}
with ${\cal R}_{\rm c}$ the cutoff rigidity. Compared with the Geminga SNR, whose injection is burst-like, a continuous injection process of electron-positron pairs is considered for the Geminga pulsar. Therefore the time-dependent injection rate  $Q_{0}(t)$ at $1$ GV is written as
\begin{equation}
Q_{0}(t) = \begin{cases}
q_{0}^{SNR} \delta(t+t_{\rm inj}^{SNR}) ~, ~ \rm for ~ SNR ~, \\

\dfrac{q_{0}^{psr} }{[1+(t+t_{\rm inj}^{psr})/\tau_0]^2} ~, ~ \rm for ~ pulsar ~,
\end{cases}
\end{equation}
where $t_{\rm inj}^{SNR/psr}$ and $\tau_0$ are the initial injection time and characteristic duration \citep{2010ApJ...710..958K, 2013PhRvD..88b3001Y}.  {The injected energy spectral index for SNR and pulsar, which are listed in Table \ref{tab:para}, are different.} The solution of Green function $G(\vec{r}, t, E)$ of the time-dependent propagation equation can be found in \citep{2004ApJ...601..340K}. The propagated spectrum from Geminga SNR and pulsar is thus a convolution of Green function with time-dependent injection rate $Q({\cal R},t)$ \citep{2010ApJ...710..958K, 2012MNRAS.421.3543K}, i.e.
\begin{equation}
\varphi(\vec{r}, {\cal R}, t) = \int_{t_{inj}}^{t} G(\vec{r}-\vec{r}^\prime, t-t^\prime, {\cal R}) Q({\cal R},t^\prime) d t^\prime .
\end{equation}

\section{Results}
The propagation parameters used in this work refer to \cite{2020ChPhC..44h5102T}, which well fit the recent observations of proton, helium and B/C ratio.  {The diffusion coefficient parameters are ${\cal D}_{0} = 9.65\times 10^{28} ~\rm cm^2 \cdot s^{-1}$  at ${\cal R}_{0}^{'}$ = 4 GV, $\delta_{0}$ = 0.65, $N_{m}$ = 0.24 and $\xi $= 0.12. The thickness of the propagation halo is $z_{h}$ = 5 kpc, and the Alfv\'en velocity is $v_{A} = 6 ~\rm km \cdot \rm s^{-1}$.} The corresponding injected spectra of proton and helium are applied to calculate the yields of secondary electrons and positrons during the transport, with the production cross section taken from the Fluka package \citep{2016APh....81...21M}. The Geminga pulsar is assumed to be $r = 0.25$ kpc away from the solar system and $t_{\rm inj}^{psr} =2\times 10^5$ years old. {At low energies the behavior of continuous injection from Geminga pulsar is similar to burstlike injection. However, at higher energies it is qualitatively different and the break-off due to the absortion is disppeared \cite{1995PhRvD..52.3265A}. }

\begin{figure*}[!htb]
	\centering
	\includegraphics[height=18cm, angle=0]{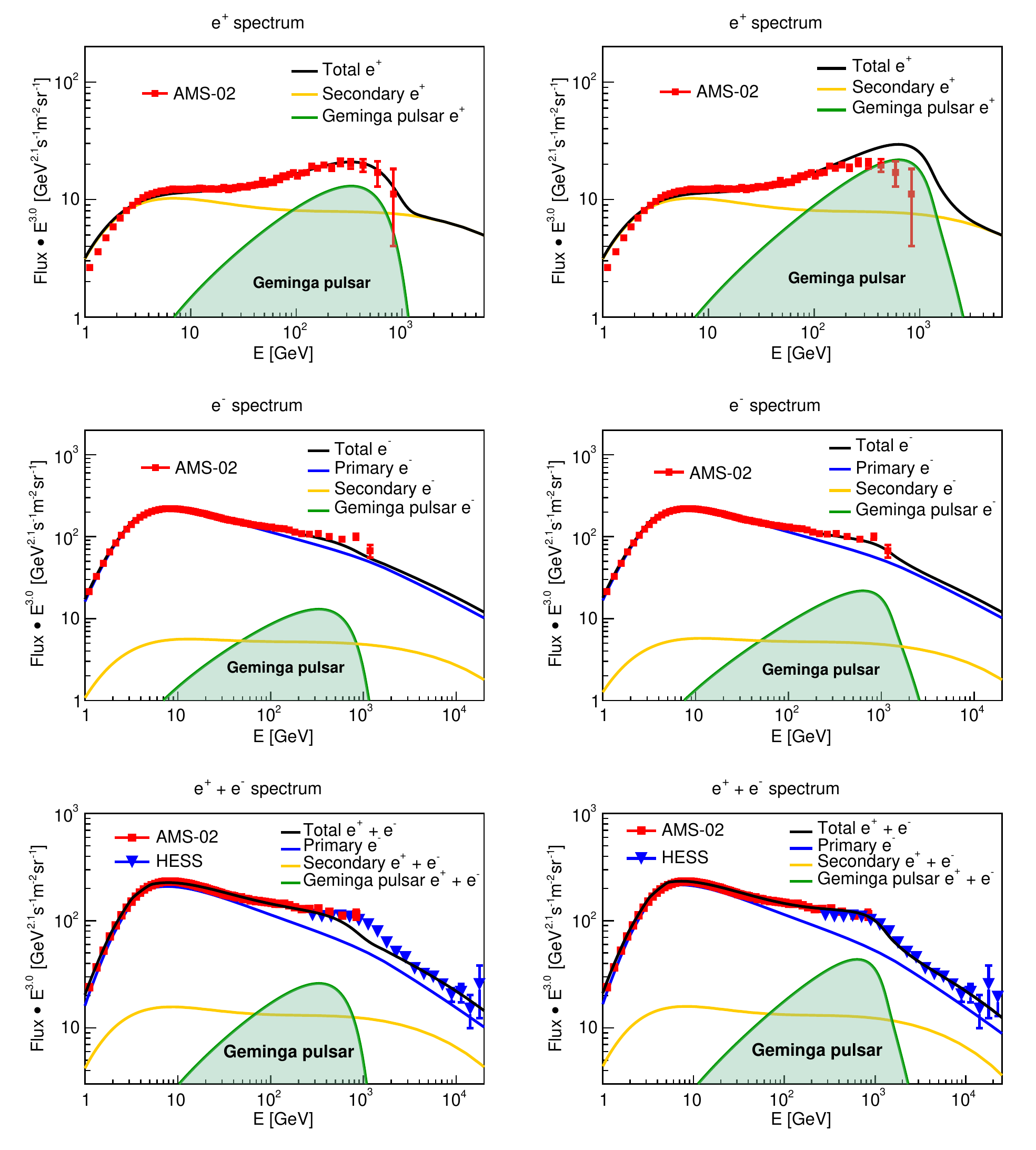}
	\caption{Calculated spectra of positrons(top), electrons(middle) and total spectrum of positron and electron(bottom). The red and blue data points are measured by AMS-02 and HESS experiments. In the left panel, ${\cal R}_{\rm c}$ of the injection spectrum of Geminga pulsar is $800$ GV, while in the right panel, ${\cal R}_{\rm c}$ is $2300$ GV. All of the propagated and injection parameters are listed in Table %\ref{tab:para}.
	}
	\label{fig:elec1}
\end{figure*}

%the steepening of this spectrum at high energies. In addition to this, the cutoff of spectrum expected in the case of burstlike injection, now disappears
Fig. \ref{fig:elec1} illustrates the calculated spectra of positrons (top), electrons (middle) and total spectrum of positron and electron (bottom). For both electrons and positrons, the measurements by the AMS-02 experiment are taken. As for total spectrum of positron and electron, the latest observation of HESS is also included as blue data points. The blue and yellow solid lines are the primary electrons from SNRs and secondary electrons/positrons generated during the propagation respectively. The green solid lines are the electrons/positrons generated by the Geminga pulsar.

%===========================================================================================================================
%===========================================================================================================================

First of all, we attempt to account for the positron excess, which is shown in the left panel of Fig. \ref{fig:elec1}. As can be seen that due to the SDP, the propagated background positron and electron spectra flatten above tens of GeV, compared with the conventional propagation model. This is similar to the spectral hardening of cosmic-ray nuclei during SDP. But unlike the CR nuclei, the energy loss of electrons shifts that excess to the lower energy, i.e. about tens of GeV. This is consistent with the hardening of electrons uncovered by the recent experiments. Meanwhile due to the FLUKA generation cross section as well as the SDP + spiral scenario, the background secondary positron flux also enhances overall greatly so that the positron spectrum above $10$ GeV could be well accounted for. These features have been entensively discussed in the previous work \citep{2020ChPhC..44h5102T}.

The calculations show that to describe the positron break at $\sim 284$ GeV, as shown in the top left panel of Fig. \ref{fig:elec1}, the cutoff rigidity $R_{\rm c}$ in the injection spectrum of the Geminga pulsar is $800$ GV. However we notice that when adopting this value, both electron and total spectrum of positron and electron are inadequate to explain above $\sim 200$ GeV. This is particularly evident in the total spectrum of positron and electron when we allow for the HESS data above $1$ TeV. The calculated total spectrum of positron and electron is far less than the measurements between $\sim 200$ GeV and $2$ TeV.

We further change the cutoff rigidity $R_{\rm c}$ to fit the total spectrum of positron and electron and study its influence of the TeV break  on the positron spectrum, as demonstrated in the right panel of Fig. \ref{fig:elec1}. In this case, ${\cal R}_{\rm c}$ is $2300$ GV in order to fit the total spectrum of positron and electron. But the positron flux continue rising above $200$ GeV until $\sim 600$ GeV then drop off. Its flux significantly exceeds the lastest AMS-02 measurements above $\sim 300$ GeV. This excess results from the higher energy cutoff of the positrons. Therefore it is obvious that if the drop-off of positron flux measured by the AMS-02 experiment is correct, inevitably there are extra electrons.

\begin{figure*}[!htb]
\centering
\includegraphics[height=5.5cm, angle=0]{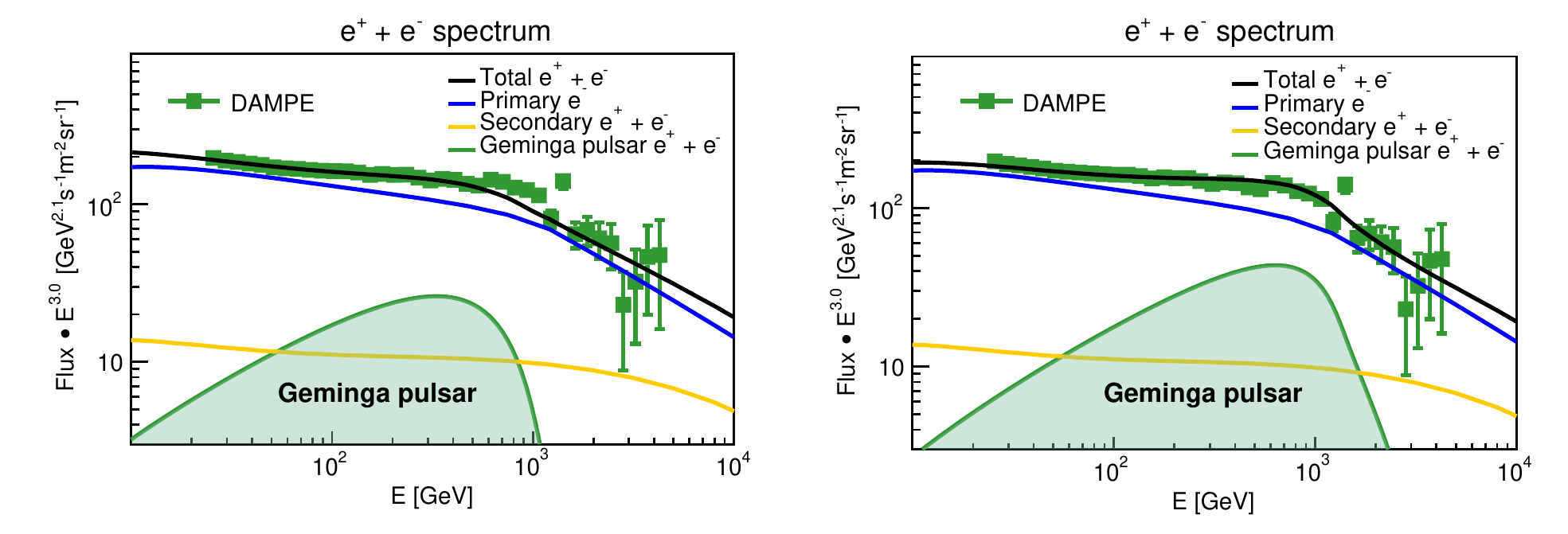}
\caption{The same calculation for DAMPE.}
\label{fig:dampe}
\end{figure*}

It should be emphasized that this excess also holds for other measurements, e.g. DAMPE. Compared with AMS-02, the DAMPE expeiment could observe CREs beyond TeV energies, from tens of GeV to $5$ TeV, well covering the energy range of our concern. But the measured total spectrum of positron and electron is overall higher than that of AMS-02 when the energies exceed $70$ GeV. Moreover, it has no measurements of electrons and positrons because it is incapable of separating positrons from electrons. Nevertheless, for the background component, the electron flux far exceeds the positron flux, therefore the background flux could be approximated as the electron contribution. The parameters of electron flux fitting DAMPE experiment are listed in Table \ref{tab:para}. Since the measured total spectrum of positron and electron by {DAMPE} is harder than AMS-02, the background electron flux is harder, with power index is $2.55$ less than $1100$ GeV and $3.0$ above that energy. It can be found that to explain the whole energy range of DAMPE experiment, the high energy cutoff of Geminga pulsar have to be $2300$, i.e. model B, as shown in the right panel of Fig. \ref{fig:dampe}. If the cutoff rigidity of model A are adopted, the similar absence also appears at around TeV energies, see the left panel of Fig. \ref{fig:dampe}.

%===========================================================================================================================
%===========================================================================================================================

The feature of excess in the total spectrum of positron and electron are further studied. After subtracting the theoretical model from the experimental data points, an additional bump of electrons appears, which is shown as circles in Fig. \ref{fig:local}. The red circles are the residuals of AMS-02 minus model, whereas the blue circles are the residuals of HESS minus model. We can see both residual are consistent. The residual electrons for DAMPE measurement are shown in the right panel of Fig. \ref{fig:local}. The blue solid lines shows the propagated electrons from Geminga SNR. As can be seen that both bumps of HESS and DAMPE are consistent and they can be well described by a time dependent propagation of local electrons. Above TeV energies, the spectrum falls off rapidly. However, the Klein Nishina effect of the inverse Compton scattering may also have some effect on regulating the electron spectrum\citep{2010ApJ...710..236S, 2020PhRvL.125e1101E, 2020arXiv200715601F}.
The injected parameters used to fit HESS are also hold for DAMPE, which are listed in Table \ref{tab:para}.

To make this work complete, the spectrum of proton is also calculated following the work \citep{2019JCAP...10..010L, 2019JCAP...12..007Q}, as shown in Fig. \ref{fig:proton}. 
%\textcolor {red}
{It is clear that the model calculation is consistent with the measurements.}
%when ${\cal D}_{0} = 9.65\times 10^{28}  \rm cm^2 s^{-1}$ at ${\cal R}_{0}^{'} = 4$ GV. }
Here the shade one is from Geminga SNR, accelerated along with the electron.
{The Geminga SNR injection spectrum is parameterized as a cutoff power-law form, with a power-law index of 2.1 and a cutoff rigidity of 14 TV. The normalization is determined through fitting proton spectrum, which results in a total energy of $\sim 3.5\times 10^{50}$ erg for protons. {If 10\% of kinetic energy is assumed to convert to accelerate CRs, the total energy of Geminga supernova explosion is estimated to $\sim3.5\times 10^{51}$ erg.}
%which is about $35\%$ of the shock kinetic energy of a typical core-collaspe supernova.}

%===========================================================================================================================
\begin{figure*}
\centering
\includegraphics[height=6cm, angle=0]{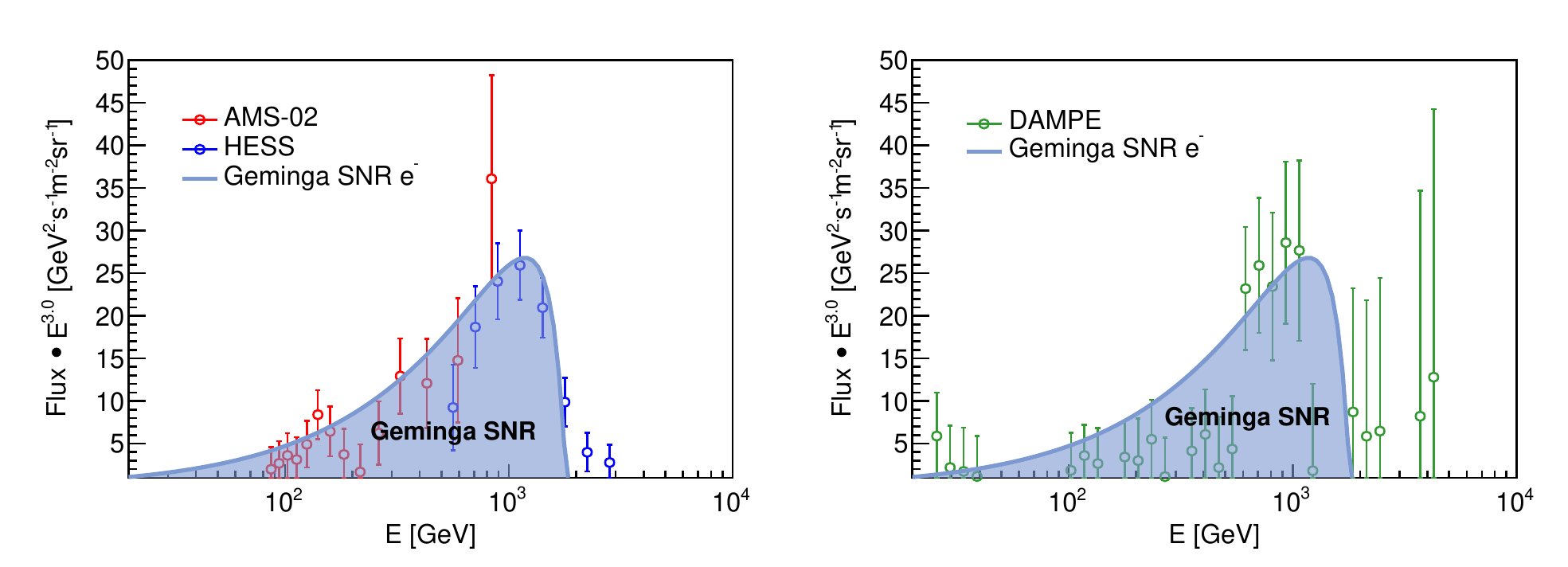}
\caption{Calculation of electrons from a Geminga SNR to explain the electron excess above $\sim200$ GeV.}
\label{fig:local}
\end{figure*}
%===========================================================================================================================

%===========================================================================================================================
\begin{figure*}[!htb]
\centering
\includegraphics[height=9cm, angle=0]{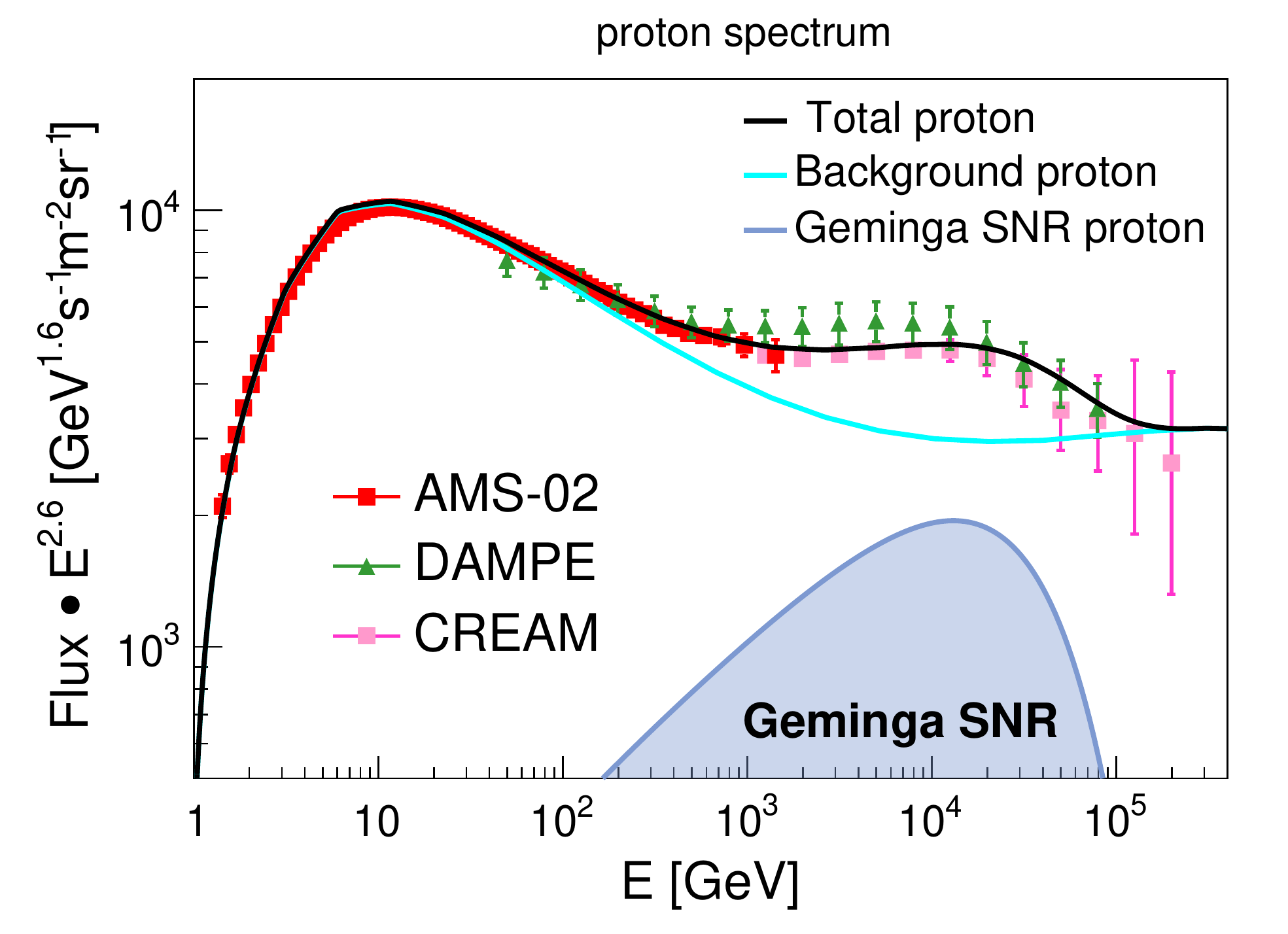}
\caption{Calculated proton spectrum.
}
\label{fig:proton}
\end{figure*}

%===========================================================================================================================
\begin{table*}
%\begin{center}
\begin{tabular}{c|c|c|c|c|c|c}
   \hline
background  &  ${Q}_0~ [\rm m^{-2}sr^{-1}s^{-1}GeV^{-1}]^\dagger$  & $\nu_{1}$  & ${\cal R}_{\rm br1}$ ~[GV] & $\nu_{2}$ & ${\cal R}_{\rm br2}$ ~[GV] & $\nu_{3}$ \\
   \hline
$e^-$ for HESS fitting    & $1.23\times 10^{-4}$ & $1.14$ & $5$ & $2.69$   & $1500$ & $2.8$\\
$e^-$ for  DAMPE fitting   & $1.35\times 10^{-4}$ & $ $ & $ $ & $2.55$ & $1100$  & $3.0$  \\
proton    & $4.22\times 10^{-2}$ & $1.9$ & $5.5$ & $2.39$   &   $10^{6}$  & $3.0$\\
\hline
Geminga pulsar &  $r_{psr}$ [kpc] & $t_{\rm inj}^{psr}$ [yrs] & $q_0^{psr}~ [\rm GeV^{-1}s^{-1}]$  & $\gamma_{psr}$  & $\tau_{0}$ [yrs]  &  ${\cal R}_{\rm c}^{\rm e_\pm}$ ~[GV]     \\
\hline
A  & $0.25$  & $2\times 10^5$ & $9.85\times 10^{49}$  & $2.25$  & $10^4$  & $800$ \\
B  & $0.25$ & $2\times 10^5$ & $1.15\times 10^{50}$ &  $2.25$  & $5\times 10^4$ & $2300$   \\

\hline
Geminga SNR &  $r_{SNR}$ [kpc] & $t_{\rm inj}^{SNR}$ [yrs] & $q_0^{SNR}~ [\rm GeV^{-1}]$  & $\gamma_{SNR}$ &  ${\cal R}_{\rm c}$ ~[GV]     \\
\hline
electron  & $0.33$ & $3.3 \times 10^5$ & $5\times 10^{49}$ &  $2.1$  & $1.4\times 10^{4}$ &    \\

proton  & $0.33$ & $3.3 \times 10^5$ & $2.2\times 10^{52}$ &  $2.1$  & $1.4\times 10^{4}$ &    \\
\hline
\end{tabular}\\
{$^\dagger${The normalization is set at reference rigidity ${\cal R}_{0} = 100$ GV.}}
%\end{center}
\caption{The injected parameters of spectra for positron, electron and proton.}
\label{tab:para}
\end{table*}
%===========================================================================================================================

%%%%%%%%%%%%%%%%%%%%%%%%%%%%%%%%%%%%%%%%%%%%%%%%%%%%%%%%%%%%%%%%%%%%%%

\section{Conclusion}

The precise measurements of CRs enable us to better reveal their origins. Recently, AMS-02 collaboration have upgraded their measurements of CR electrons and positrons. Compared with the previous observations, a new positron break-off above $\sim 284$ GeV has been found. This is different from the electron spectrum, whose power-law spectrum extends to the higher energy and has a break at about TeV energies. In this work, we study the high energy cutoff of electron and positron spectra based on the SDP model. A conventional SNR background plus local pulsar model is applied. We find there is a tension in the high energy cutoff between the positron and total spectrum of positron and electron in this simple scenario. Compared with total spectrum of positron and electron, which requires the cutoff rigidity of Geminga puslar is $2300$ GV, to well fit positron dropoff, the high energy cutoff of the Geminga pulsar is just $800$ GV. Therefore if the observed cutoff of positron flux is correct, there are extra electrons from $\sim 200$ GeV to $2$ TeV which are  not accounted for. This excess of electrons has not be been claimed as far as we know.

After subtracting the calculated electron and positron fluxes from the measured total spectrum of positron and electron, the excessive electrons has a bump-like structure. And we show this excess is common whatever AMS-02, HESS or DAMPE measurements. This feature indicates an origin of local sources. One of probable source is Geminga SNR. This source has been used to account for both spectral hardening of CR nuclei and large-scale anisotropy. Meanwhile the Geminga pulsar is widely believed to be source of the positron excess. We find that the extra component could be decently explained by the Geminga SNR. Therefore the observational features of Galactic cosmic rays could be accounted for in our unified picture with SNR background plus a local source, even this excess of electons. At present, the measurements of positrons above $200$ GeV are still inaccurate, with quite large errors. We expect the coming precise measurements could testify our speculation.
%%%%%%%%%%%%%%%%%%%%%%%%%%%%%%%%%%%%%%%%%%%%%%%%%%%%%%%%%%%%%%%%%%%%%%
\acknowledgments
This work is supported by the National Key R\&D Program of China grant No. 2018YFA0404203, the National Natural Science Foundation of China (Nos. 11635011, 11875264, U1831129, 11722328, 11851305, U1738205, U2031110).

\bibliographystyle{unsrt_update}
\bibliography{ref1}

\end{document}